# Physical Bounds on Absorption and Scattering for Cloaked Sensors

Romain Fleury, Jason Soric and Andrea Alù[*]

Dept. of Electrical and Computer Engineering, The University of Texas at Austin

1 University Station C0803, Austin, TX 78712, USA

*alu@mail.utexas.edu

*We derive and discuss general physical bounds on the electromagnetic scattering and absorption of passive structures. Our theory, based on passivity and power conservation, quantifies the minimum and maximum allowed scattering for an object that absorbs a given level of power. We show that there is a fundamental trade-off between absorption and overall scattering suppression for each scattering harmonic, providing a tool to quantify the performance of furtive sensors, regardless of the applied principle for scattering suppression. We illustrate these fundamental limitations with examples of light scattering from absorbing plasmonic nanoparticles and loaded dipole antennas, envisioning applications to the design of cloaked sensors and absorbers with maximized absorption efficiency.*

I. INTRODUCTION

In the past decade, the successful design and implementation of metamaterials with exotic electromagnetic properties has opened several new venues for manipulating the propagation of light and its interaction with matter [1]. In particular, the possibility of inducing invisibility with passive metamaterial coatings, the "cloaking effect", has been the subject of intense research [2-19]. Several experimentally-verified approaches for all-angle electromagnetic cloaking of three-dimensional objects have been proposed, including (i) the transformation method [5-8], which uses a cloak with functionally graded material properties to re-route the power flow around the object, simultaneously bringing its scattering to zero and insulating the cloaked region from the outside world; (ii) the scattering cancellation technique [9-19], in which a metamaterial coating with isotropic and homogeneous constitutive parameters [9-16], or a thin metasurface with tailored surface impedance [17-19], is used to reduce the scattering over a given bandwidth. In many cases, the study of the cloaking performance has been restricted to ideal situations involving lossless materials. However, the problem of material losses is a central issue, especially in practical realizations, in which the absolute cloaking reduction may be significantly affected by absorption losses [3,20].

A related problem that has recently attracted significant attention is the one of "seeing without being seen" [21,22]. In many applications like non-invasive sensing and communications, being able to "open our eyes" behind a cloak, and extract information about the outside world while remaining undetectable, is of primary importance. This possibility has been demonstrated with the scattering cancellation technique and the transformation approach in the case of small sensors or power receivers [23-28], providing exciting venues in a variety of application fields like near-field scanning optical microscopy [29-31], invisible electromagnetic sensors and photodetectors [32] and low-observable receiving antennas [33]. Yet, the idea of being invisible while absorbing power appears counterintuitive, since common sense and the optical theorem agree on the fact that it is not possible to absorb energy without creating a shadow, i.e., any extinction is associated with nonzero forward scattering [34,35]. As a consequence, a passive cloaked object becomes necessarily detectable as soon as it starts to absorb a portion of the impinging energy.

A better understanding of the fundamental limitations associated with cloaking absorptive objects is not only crucial to transition from ideal cloaking methods to practical applications, but also for understanding the physical boundaries to be considered when building concealed sensors. In this article, we discuss general limitations on scattering and absorption from passive objects, stemming from passivity and power conservation. In Section II, we highlight and quantify the fundamental trade-off between absorption and cloaking, and provide an analytical tool to understand the behavior of cloaked sensors and low-interfering power receivers. In Section III, we illustrate the generality of our theory by considering practical examples, including optical scattering from lossy nanospheres, core-shell nanoparticles and loaded dipole antennas.

II. THEORETICAL FORMULATION

Consider the general situation in which electromagnetic waves are scattered off from a passive object. For simplicity of notation, in the following we assume spherical symmetry, but our theory can also be extended to arbitrarily shaped objects. The scattering problem may be approached using the Mie expansion in spherical harmonics. The scattered field for plane wave incidence $\vec{E}_{inc} = \hat{x} E_0 e^{ik_0 z}$ is expressed as a superposition of spherical harmonics [34]

$$\vec{E}_{scat} = E_0 \left( \sum_{n=1}^{+\infty} c_n^{TM} \nabla \times \nabla \times \left(r\psi_n^1\right) + i\omega\mu_0 \sum_{n=1}^{+\infty} c_n^{TE} \nabla \times \left(r\psi_n^1\right) \right), \tag{1}$$

where $\mu_0$ is the free-space permeability, $\psi_n^m$ are scalar spherical harmonics, solutions of the Helmholtz equation in the spherical coordinate system $(r,\theta,\varphi)$, and $m=1$ due to symmetry, under an $e^{-i\omega t}$ time convention. The total scattering cross-section $\sigma_{scat}$ can be expressed as a function of the Mie scattering coefficients $c_n^{TE}$ and $c_n^{TM}$ as [34,35]

$$\sigma_{scat} = \frac{2\pi}{k_0^2} \sum_{n=1}^{+\infty} (2n+1) \left[ \left|c_n^{TM}\right|^2 + \left|c_n^{TE}\right|^2 \right], \tag{2}$$

where the scattering coefficients may be found for a general core-shell geometry in [36]. The total amount of power extracted from the incident field is represented by the total extinction cross-section $\sigma_{ext}$, which may be calculated as [34,35]

$$\sigma_{ext} = -\frac{2\pi}{k_0^2} \sum_{n=1}^{+\infty} (2n+1) \operatorname{Re}\left[ c_n^{TM} + c_n^{TE} \right]. \tag{3}$$

In addition, the absorption $\sigma_{abs}$, extinction and scattering cross-sections are related by energy conservation

$$\sigma_{abs} = \sigma_{ext} - \sigma_{scat}. \tag{4}$$

We will now prove that, as a consequence of passivity ($\sigma_{abs} \geq 0$), the complex-valued scattering coefficients $c_n^{TE}$ and $c_n^{TM}$ are restricted to a portion of the complex plane, namely the closed disk of center $-1/2$ and radius $1/2$. This will be evident after performing the variable change

$$c_n^{TE/TM} = -\frac{1}{2} + \frac{1}{2}\eta_n^{TE/TM}. \tag{5}$$

The expressions (2)-(4) for the cross-sections become:

$$\sigma_{scat} = \sum_{n=1}^{+\infty}(\sigma_{scat})_n = \frac{\pi}{2k_0^2}\sum_{n=1}^{+\infty}(2n+1)\left[\left|\eta_n^{TM}-1\right|^2 + \left|\eta_n^{TE}-1\right|^2\right], \tag{6}$$

$$\sigma_{ext} = \sum_{n=1}^{+\infty}(\sigma_{ext})_n = \frac{\pi}{k_0^2}\sum_{n=1}^{+\infty}(2n+1)\left[1-\operatorname{Re}\eta_n^{TM}+1-\operatorname{Re}\eta_n^{TE}\right], \tag{7}$$

$$\sigma_{abs} = \sum_{n=1}^{+\infty}(\sigma_{abs})_n = \frac{\pi}{2k_0^2}\sum_{n=1}^{+\infty}(2n+1)\left(1-\left|\eta_n^{TE}\right|^2+1-\left|\eta_n^{TM}\right|^2\right), \tag{8}$$

where we have introduced the partial scattering cross-sections associated with each harmonic, $(\sigma_{scat})_n$, $(\sigma_{abs})_n$ and $(\sigma_{ext})_n$. Inspecting (6), it is evident that a perfectly cloaked object ($\sigma_{scat}=0$) requires $\eta_n^{TE/TM}=1$, which implies $\sigma_{abs}=0$ in (8). In addition, if the scatterer is lossless ($\sigma_{abs}=0$) then necessarily $\left|\eta_n^{TE/TM}\right|=1$ $\forall n$. If losses are present, passivity and the orthogonality of spherical harmonics require that $(\sigma_{abs})_n \geq 0$, which translates into $\left|\eta_n^{TE/TM}\right| \leq 1$. This proves that, for passive objects, the frequency dependent complex coefficients $\eta_n^{TE/TM}$ are restricted in the complex plane to the closed unity disk, or equivalently, the coefficients $c_n^{TE/TM}$ are restricted to the closed disk of center $-1/2$ and radius $1/2$. A special case of interest is the one of maximized absorption for the $n^{th}$ harmonic. Equation (8) suggests that $(\sigma_{abs})_n$ is maximized for $\eta_n^{TE/TM}=0$. Under this condition, we find the relationship between partial cross-sections

$$\left(\sigma_{abs}\right)_n = \left(\sigma_{scat}\right)_n = \frac{\left(\sigma_{ext}\right)_n}{2} = \left(2n+1\right)\frac{\pi}{k_0^2}. \tag{9}$$

In other words, in the case of maximized absorption for the $n^{th}$ harmonic, the partial absorption and scattering cross-sections are necessarily equal and they depend only on frequency and on the order $n$, a condition known in the antenna community as 'conjugate matched' resonance [37-41].

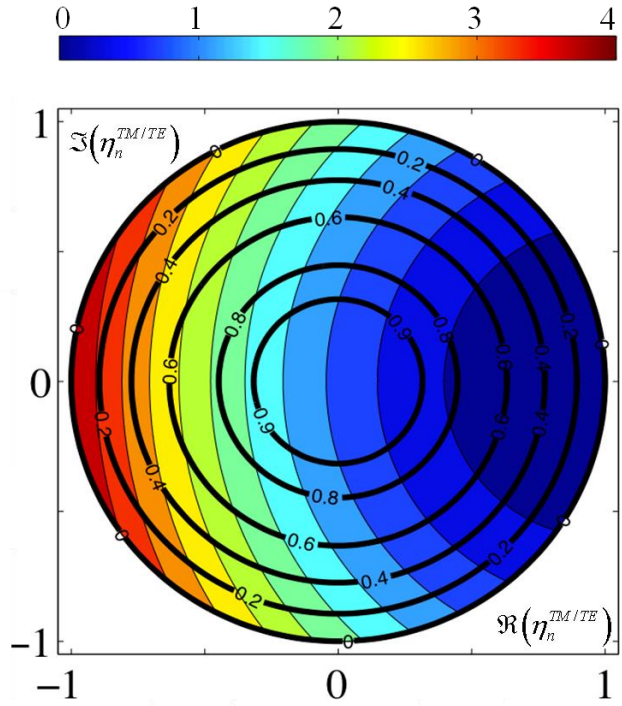

Figure 1. Partial scattering cross-sections for passive objects, represented inside the unity disk in the $\eta$ complex plane. The filled contours represent the levels of $\left(\sigma_{scat}\right)_n$, while the black contour lines represent $\left(\sigma_{abs}\right)_n$. Both are expressed in normalized units of $(2n+1)\pi/2k_0^2$.

The above discussion demonstrates that the scattering and absorption cross-sections of any passive object are bounded and fundamentally related. To further illustrate this concept, we plot in Figure 1 the partial absorption and scattering cross-sections of an arbitrary passive object as a function of $\eta_n^{TE/TM}$ in the closed unity disk of the complex plane. The scattering cross-section is shown in the filled contour plot, while absorption is represented with black contour lines. Both partial cross-sections are expressed in normalized units of $(2n+1)\pi/2k_0^2$, so that the plot remains valid for any order $n$. This figure is a powerful tool, as it includes all the available information on absorption, scattering and extinction for each harmonic. As expected, extreme values of the scattering cross-section are obtained on the unity circle, for $\eta = \pm 1$, for which the absorption is zero. This demonstrates the importance of minimizing losses when maximal (strong scattering resonance) or minimal (cloaking) scattering is desired. Cloaked sensors lie on the real axis in the range $\text{Re}\,\eta \in [0,1]$, in which the ratio between absorption and scattering is maximized for a given absorption level, as it can be inferred from the figure.

We define now the *absorption efficiency* for the $n$-th harmonic as the ratio $(\sigma_{abs})_n / (\sigma_{scat})_n$. Combining (8) and (6), this quantity may be generally written as

$$\left(\frac{\sigma_{abs}}{\sigma_{scat}}\right)_n^{TE/TM} = \frac{1 - \left|\eta_n^{TE/TM}\right|^2}{\left|\eta_n^{TE/TM} - 1\right|^2}. \tag{10}$$

For electrically small objects dominated by dipolar scattering, a case of particular interest in the case of cloaked sensors [21-28], this quantity for $n=1$ coincides with the total absorption efficiency $\sigma_{abs}/\sigma_{scat}$. In principle this quantity can be made as large as possible, but, because of passivity, $\left|\eta_n^{TE/TM}\right| \leq 1$ and the absorption efficiency is fundamentally bounded by the total amount of absorption, as we show in Figure 2, in which we calculate $\left(\sigma_{abs}\right)_n / \left(\sigma_{scat}\right)_n^{TE/TM}$ and $\left(\sigma_{abs}\right)_n^{TE/TM}$ for every admissible value of $\eta_n^{TE/TM}$, essentially drawing the image of the unity disk of Fig. 1 into the $\left(\sigma_{abs}\right)_n / \left(\sigma_{scat}\right)_n^{TE/TM}$ versus $\left(\sigma_{abs}\right)_n^{TE/TM} / (2n+1)\lambda_0^2$ plane, corresponding to the blue hachured region in Figure 2. This area is interestingly limited by a fundamental physical bound (solid black line) for the absorption efficiency of each harmonic as a function of the absorption level. This is consistent with our recent findings for inelastic quantum scattering, derived in the context of designing cloaked sensors for matter waves [42].

Eq. (10) ensures that the boundary of the admissible region (black solid line) is obtained for real values $-1 \leq \eta_n^{TE/TM} \leq 1$. If absorption is maximized for any harmonic, the efficiency is 1, consistent with (9), and the normalized absorption is $1/(8\pi)$, consistent with the most right point on the black line in Fig. 2. From this point, it is possible to either decrease or increase the absorption efficiency, and bring it to any arbitrarily large level of interest, but only at the cost of sacrificing part of the absorption. For sufficiently low absorption, it is possible to significantly suppress the scattering, as quantified in Figure 2. This physical bound described here is of primary importance for the design of passive cloaked sensors, as it highlights the ultimate limits of performance for each scattering harmonic, and how close to the optimal scattering reduction we are for a given level of partial absorption coefficient.

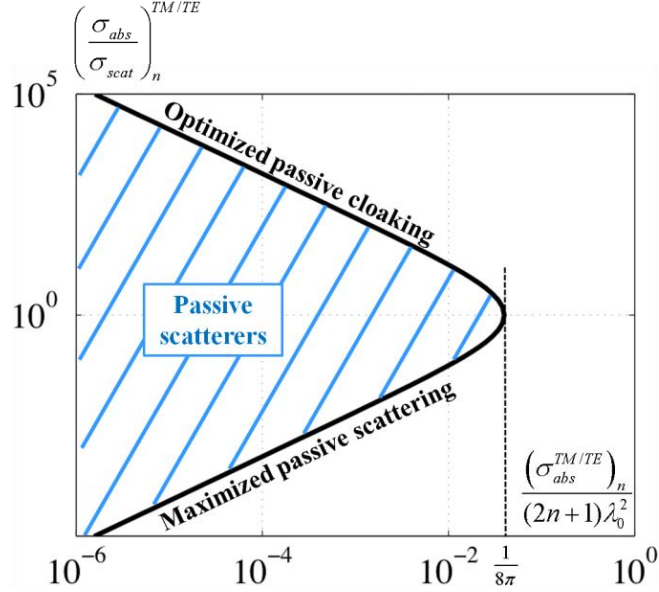

Figure 2. Scattering bound for absorptive, passive objects, valid for any scattering harmonic order $n$. Only the blue shaded region on this plane is admissible, yielding an ultimate limit on the minimum scattering required for a given level of absorption.

## III. EXAMPLES AND DISCUSSION

In this section, we apply the theory developed in Section II to concrete scattering examples. Our goal is to underline the generality of our theory, discuss its implications and show how it may be viewed as an essential tool to understand and optimize invisible absorbers and sensors.

### A. Optical scattering from lossy nanospheres

As a first example, we consider the scattering from nonmagnetic nanospheres in the presence of material losses. We show that this simple passive system complies with our general bounds, and we use our formalism to study and better explain the general relation between scattering and absorption in this example. This scattering problem is solved in detail in the Appendix. A central question of interest is whether or not this passive system can reach the ultimate bounds derived in Figure 2, and under what conditions. In order to approach this question analytically, we first assume that the nanosphere is electrically small, of size $x = k_0 a \ll 1$, $a$ being the sphere radius. Under this assumption, the scattering coefficients $c_n^{TM}$ obtained from (A8) reduce to the quasistatic expression

$$c_n^{TM} = -\left(1 + i\frac{4^n(2n+1)\Gamma(n+1/2)^2}{\pi(n+1)x^{2n+1}}\frac{(\varepsilon n + n + 1)}{\varepsilon - 1}\right)^{-1}. \tag{11}$$

According to the previous section, in order to lie on the boundary of the allowed region for the *n*-th TM harmonic, and therefore achieve the most interesting scattering properties for the given level of absorption, one needs to have $\mathrm{Im}(\eta_n^{TM}) = \mathrm{Im}(2c_n^{TM} + 1) = 0$. Using Eq. (11), we find that this condition is met whenever the permittivity $\varepsilon = \varepsilon' + i\varepsilon''$ of the nanosphere satisfies

$$\varepsilon'_\pm = \frac{-1 \pm \sqrt{1 + 4n(1 + n - n\varepsilon''^2)}}{2n}. \tag{12}$$

Assuming that we are able to tailor the sphere permittivity at will, for a fixed level of losses $\varepsilon'' > 0$ it is possible to reach the fundamental bound for two distinct values of $\varepsilon'$ as long as

$$0 \leq \varepsilon'' < 1 + \frac{1}{2n}. \tag{13}$$

In the particular case of small losses $\varepsilon'' \ll 1$, the plus sign solution in (12) simplifies into

$$\varepsilon'_+ = \varepsilon'_{transparency} = 1 - \frac{n\varepsilon''^2}{2n+1} + o(\varepsilon''^3). \tag{14}$$

In the limit $\varepsilon'' \to 0^+$, this solution simply converges to the transparency condition that minimizes the scattering from the sphere, i.e., $\varepsilon \approx 1 + o(\varepsilon''^2)$. When small losses are present, condition (14) ensures that we hit the bound on the upper portion of the curve, maximizing the absorption efficiency for the given level of absorption. It is worth mentioning that this condition is in general not identical to the solution that minimizes the scattering for the chosen level of $\varepsilon''$, as it differs from it by a quantity proportional to $o(\varepsilon''^2)$.

Conversely, in the same low-loss limit the minus sign solution of Eq. (12) simplifies into

$$\varepsilon'_- = \varepsilon'_{resonance} = -\frac{n+1}{n} + \frac{n\varepsilon''^2}{2n+1} + o(\varepsilon''^3). \qquad (15)$$

In the lossless limit, this solution converges to the condition that maximizes the scattering cross section of the sphere, as it coincides with the plasmonic resonance $\varepsilon' = -(n+1)/n$ [43], and in presence of small losses condition (15) allows again hitting the bound. Also in this case this condition differs from the condition to maximize absorption for the given level of $\varepsilon''$ by a second-order term in $\varepsilon''$. The two conditions derived above represent the required values of $\varepsilon'$, for a given level of $\varepsilon''$, to reach the solid black boundary in Fig. 2. Note that it is obviously always possible to find a suitable value of $\varepsilon'$ that maximizes the absorption or minimizes the scattering for the chosen value of $\varepsilon''$, but only in the limit $\varepsilon'' \to 0^+$ these solutions lie on the bound of Fig. 2, according to Eqs. (14)-(15).

The above quasi-static analysis is important to unveil the complexity of the scattering problem in relation to our physical bounds in the general dynamic case. To validate our findings, we have numerically calculated absorption and scattering in the fully dynamic case for a nanoparticle of small electrical size $x = 0.2$, for different values of permittivity. In Figure 3, we show the evolution of these quantities in the absorption efficiency vs. absorption plane, comparing the obtained results to the TM$_1$ physical bound represented by the black solid line. Let us first focus on the solid lines, which represent the contours obtained when sweeping $\varepsilon'$ over all real values, while keeping $\varepsilon''$ constant. The blue solid line is obtained for a relatively low value of material losses, $\varepsilon'' = 0.05$. This value of $\varepsilon''$ is significantly smaller than the critical value $3/2$, obtained by evaluating (13) for $n = 1$, value beyond which the bound cannot be reached. Therefore we expect the contour to hit the fundamental bound at two distinct points, under the resonance and the transparency conditions. Because we are in the low-loss limit, the resonance point corresponds to maximum absorption for the given value of $\varepsilon''$, and the transparency point corresponds to the maximum absorption efficiency. Indeed, the solid blue line starts in the bottom left corner for large negative values of $\varepsilon'$ and, as $\varepsilon'$ grows and get closer to -2, $\sigma_{abs}$ increases until it reaches the maximum, obtained for $\varepsilon' = -2.1$ in good agreement with the quasistatic prediction (15). This maximum lies on the physical bound (solid black line), consistent with the predictions of our quasi-static model. As we keep increasing $\varepsilon'$, $\sigma_{abs}$ monotonically decreases, but yet the absorption efficiency $\sigma_{abs} / \sigma_{scat}$ reaches a maximum on the upper portion of the bound. The associated scattering minimum is indeed obtained for $\varepsilon' \simeq 1$, consistent with (14).

When the losses are equal to the critical value (13), we obtain the red solid line in Fig. 3. Consistent with Eq. (12), in this case the solutions $\varepsilon'_{\pm}$ are now degenerate, therefore we expect the curve to be tangential to the bound, reaching it at the value $\varepsilon' = 1/2 = (\varepsilon'_+ + \varepsilon'_-)/2$. Because the lossless limit assumption is no longer valid in this case, we should not expect the maxima of absorption or absorption efficiency to occur on the bound. Indeed, these predictions are all verified in the dynamic evolution of the red contour, for which the absorption and absorption efficiency maxima occur inside the allowed region, away from the bound. By choosing $\varepsilon''$ be larger than this critical value, as in the case of the solid green line ($\varepsilon'' = 3$), we do not hit the physical bound for any value of $\varepsilon'$.

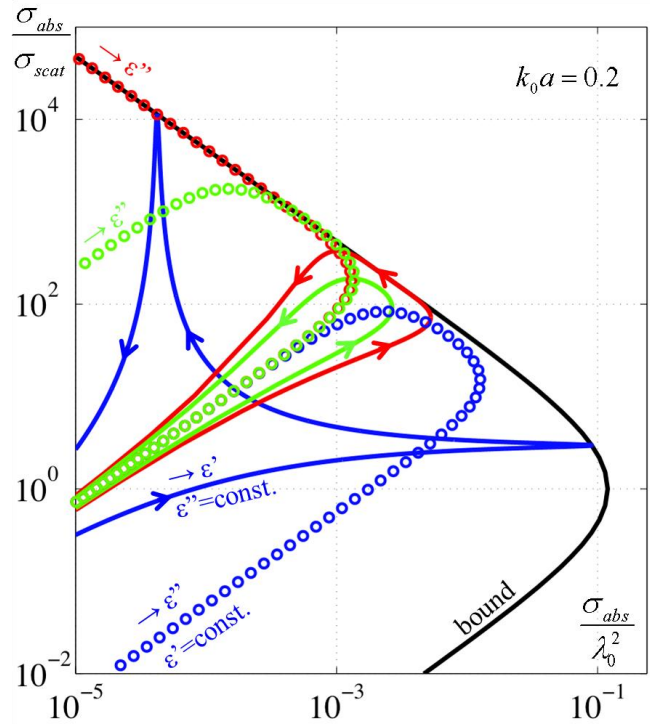

Figure 3 : Scattering and absorption for a dielectric nanosphere of permittivity $\varepsilon' + i\varepsilon''$ and electric size $x = 0.2$, for different scenarios. The black solid lines represent the $TM_1$ bound. The other solid lines represent contours obtained when sweeping $\varepsilon'$ over all real values, for $\varepsilon'' = 0.05$ (below critical, blue line), $\varepsilon'' = 1.5$ (critical, red line) and $\varepsilon'' = 3$ (beyond critical, green line). The circular markers represent the dual contours obtained by sweeping $\varepsilon''$ over all positive values, for a given value of $\varepsilon' = -2.4$ (blue markers), $\varepsilon' = 1$ (red markers) and $\varepsilon' = 0.84$ (green markers).

In Figure 3, we also plot dual contours, represented by circular markers, which correspond to the case in which $\varepsilon'$ is fixed and $\varepsilon''$ is varied over all positive numbers. The blue markers correspond to $\varepsilon' = -2.4$, the red markers to $\varepsilon' = 1$ and the green ones to $\varepsilon' = 0.84$. We notice that some of these contours appear to hit the bound, while others stay away from it. To understand this behavior, we invert condition (12) to express the condition to intersect the bound for constant $\varepsilon'$ contours, and find the unique solution

$$\varepsilon'' = \frac{1}{2}\sqrt{9 - (2\varepsilon' + 1)^2} \ . \tag{16}$$

Such a solution exists only for $-2 \leq \varepsilon' \leq 1$, i.e., for values of $\varepsilon'$ between the lossless resonance and transparency conditions. This is indeed verified in Figure 3, for which the $\varepsilon' = -2.4$ (blue) contour never crosses the bound, while the $\varepsilon' = 0.84$ does. The $\varepsilon' = 1$ contour is a limiting case, which does cross the bound only at infinity in the top left corner of the figure, for which $\varepsilon'' = 0^+$, consistent with (16). Interestingly, the $\varepsilon' = 1$ contour asymptotically converges to the bound for small values of $\varepsilon''$, but gets away from the bound for sufficiently large values of $\varepsilon''$, converging asymptotically towards the bottom left corner of the figure. In fact, all the constant-$\varepsilon'$ contours appear to tend to the same oblique asymptote when losses are sufficiently high. The constant $\varepsilon''$ contours also tend to the same asymptote when the real part of permittivity is largely positive or negative. This is easily understood by rewriting Eq. (10) for the TM$_1$ harmonic as

$$\frac{\sigma_{abs}}{\sigma_{scat}} = \frac{8\pi}{3\left|\eta_1^{TM} - 1\right|^2}\left(\frac{\sigma_{abs}}{\lambda_0^2}\right). \tag{17}$$

Due to the fact that $\eta_1^{TM}$ is bounded and well-behaved, for $\varepsilon \to \infty$ (positive, negative or imaginary) its value asymptotically converges to a constant and Eq. (17) maps into a straight line on the plane of Fig. 3. In the quasi-static limit, Eq. (17) can be explicitly written as

$$\frac{\sigma_{abs}}{\sigma_{scat}}(\varepsilon,\varepsilon') = \frac{3\pi}{2x^6} \frac{\varepsilon''^2 + (2+\varepsilon')^2}{\varepsilon''^2 + (\varepsilon'-1)^2} \left(\frac{\sigma_{abs}}{\lambda_0^2}\right)(\varepsilon,\varepsilon'), \tag{18}$$

which explicitly shows that, in the limit $\varepsilon'' \to +\infty$ or $\varepsilon' \to \pm\infty$, we always get the same linear relationship between $\sigma_{abs}/\sigma_{scat}$ and $\sigma_{abs}/\lambda_0^2$:

$$\frac{\sigma_{abs}}{\sigma_{scat}} \Big/ \left(\frac{\sigma_{abs}}{\lambda_0^2}\right) \to \frac{3\pi}{2x^6} \quad [\varepsilon'' \to +\infty \text{ or } \varepsilon' \to \pm\infty]. \tag{19}$$

As a result, on the logarithmic scale of Figure 3, all the contours converge in these limits to the asymptote

$$\log \frac{\sigma_{abs}}{\sigma_{scat}} = \log \frac{3\pi}{2} - 6\log x + \log \frac{\sigma_{abs}}{\lambda_0^2}, \tag{20}$$

for which the slope is independent of the electrical size of the nanosphere, and the intercept on the vertical axis shifts up for smaller objects. In the following we refer to this asymptote as the perfect electric conductor (PEC) limit, for obvious reasons.

In the opposite limit of zero material losses, the constant $\varepsilon'$ contours also converge to straight lines in Figure 3, to which we refer as lossless asymptotes. Their position depend on the particular value of $\varepsilon'$, but they are all parallel to the PEC asymptote. This is explained using Eq. (18) for $\varepsilon'' \to 0$. We obtain

$$\left.\frac{\sigma_{abs}}{\sigma_{scat}}\middle/\left(\frac{\sigma_{abs}}{\lambda_0^2}\right)\to\frac{3\pi}{2x^6}\left(\frac{2+\varepsilon'}{\varepsilon'-1}\right)^2\quad[\varepsilon''\to 0],\right.\tag{21}$$

so that the expression for the lossless asymptotes is given by

$$\log\frac{\sigma_{abs}}{\sigma_{scat}}=\log\frac{3\pi}{2}-6\log x+\log\left(\frac{2+\varepsilon'}{\varepsilon'-1}\right)^2+\log\frac{\sigma_{abs}}{\lambda_0^2},\tag{22}$$

indeed describing lines parallel to the PEC asymptote and shifted by an amount that depends on $\varepsilon'$. From (22), we recognize that the lossless asymptotes are shifted up with respect to the PEC asymptote when $\varepsilon'>(\varepsilon'_+ +\varepsilon'_-)/2$ and down in the opposite case. Note that the value $(\varepsilon'_+ +\varepsilon'_-)/2$ equals 0.5 for the $TM_1$ harmonic and it is independent of $\varepsilon''$, cf. (12). It is exactly the average of the resonance and transparency solutions, regardless of $\varepsilon''$. It turns out therefore that the PEC asymptote separates the transparency region (defined by $\varepsilon'>(\varepsilon'_+ +\varepsilon'_-)/2$) from the resonance region ($\varepsilon'<(\varepsilon'_+ +\varepsilon'_-)/2$). When the electrical size of the sphere varies, the PEC asymptote is accordingly shifted, dragging with it the lossless asymptotes, and therefore, all contour lines. All these findings allows us to fully describe the dynamics of the curves in Fig. 2 for arbitrary values of $\varepsilon'$ or $\varepsilon''$.

Figure 4 shows a more complete diagram for the same size sphere, considering many different values of $\varepsilon'$ and $\varepsilon''$, while focusing on low values of $\varepsilon''$ for which the bound can be reached. Similar to the previous plot, the solid lines are $\varepsilon''$-constant lines, sweeping $\varepsilon'$ from negative to positive values. The dashed contour lines are instead $\varepsilon'$-constant lines, sweeping $\varepsilon''$ through positive values. The contours essentially form a curvilinear reference system mapping an arbitrary value of complex permittivity into the corresponding level of absorption and absorption efficiency, and they are found to span the entire admissible region of Fig. 4, implying that, given the opportunity to arbitrarily vary $\varepsilon'$ and $\varepsilon''$ at the frequency of interest, we may realize any allowed level of absorption and absorption efficiency with just a single, dielectric nanoparticle. Each contour line in the figure, for a given value of $\varepsilon'$ and $\varepsilon''$ as indicated in the plot, is formed by segments of different color, and the solid and dashed curves intersect only when they have the same color.

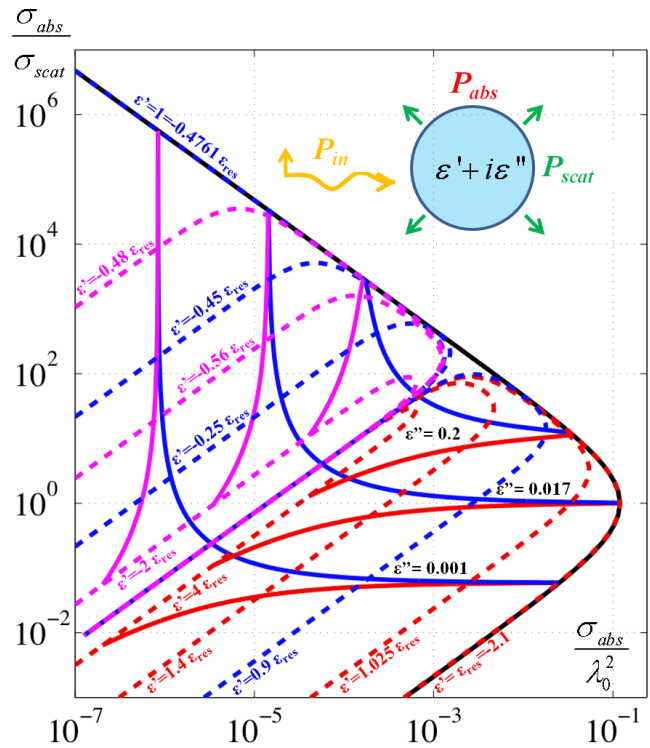

Figure 4. Similar to Fig. 3, scattering and absorption for a nanoparticle of electric size $k_0 a = 0.2$, varying its permittivity $\varepsilon' + i\varepsilon''$. The dashed contours are obtained varying $\varepsilon''$, keeping $\varepsilon'$ constant at the indicated value. The solid contours are plotted varying $\varepsilon'$, keeping $\varepsilon''$ constant at the indicated value (in black). The black solid line represents the $TM_1$ fundamental bound. Scattering and absorption for an arbitrary value of complex permittivity are obtained at the intersections of curves with same color.

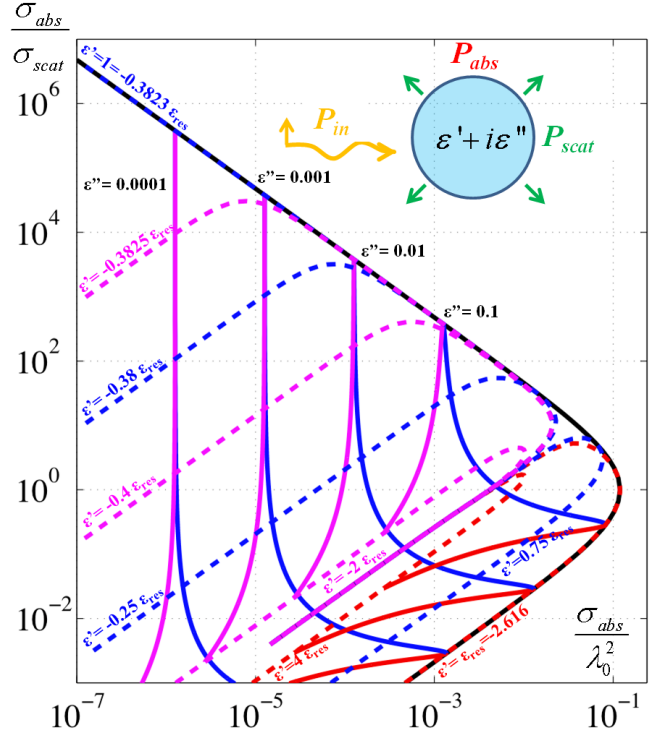

Figure 5. Same as Figure 4 but for a larger electrical size, $k_0 a = 0.5$. As predicted by the theory, the PEC asymptote is shifted down.

In Figure 5 we plot the same contours as in Figure 4, but for a bigger electrical size ($x = 0.5$), confirming the expected downward shift of the PEC asymptote consistent with (20), which carries along all the other contours. These figures synthetically confirm that the above considerations, including the asymptotic behavior, hold in the general dynamic case. Comparison with the physical bound unveils the complexity of the relation between scattering and absorption of a nanosphere, outlining the role of material losses, size and permittivity. As a corollary of our findings, it is in principle not necessary to use a cloak to obtain a high level of absorption efficiency, provided that we can arbitrarily vary $\varepsilon$, as any point on the solid black line is accessible. It would be indeed sufficient to have $\varepsilon' \simeq 1$ to enable the maximum possible absorption efficiency for a given (sufficiently low) level of $\varepsilon''$, at the transparency condition (14). Obviously in a realistic scenario we do not have the arbitrary control on the complex permittivity at the frequency of interest, and therefore a suitably designed cover may tailor scattering and absorption with more degrees of freedom, as we discuss next.

*B. Furtive optical sensors and absorbers made of core-shell nanoparticles*

The addition of a cloak is useful to change the dynamics of the scatterer response in the previous plots, when, as it is usually the case, we do not have the flexibility of changing the permittivity, the material losses, or the size at will. In Figure 6 we consider a core-shell geometry, as in the plasmonic cloaking technique [9], in which a lossless shell with radius $a_c$ and permittivity $\varepsilon_c = 0.105$ covers a core of radius $a$ with permittivity $\varepsilon = \varepsilon' + i\varepsilon''$, varied over all admissible values. The electrical size of the object is $k_0 a_c = 0.2$, for a fixed filling ratio $\gamma = a/a_c = 0.9$. The shell permittivity was chosen to reduce the scattering of a PEC sphere of same size, using the scattering cancellation method [9]. As visible in the figure, the effect of such cloak design is indeed to shift up the PEC oblique asymptote, dragging all the contour lines with it. Consistent with Eq. (20) and the nature of the proposed cloak, the effect is equivalent to reducing the effective size of the object in the PEC limit and, as a noticeable consequence, the resonance region (below the PEC asymptote) now includes a significant portion of the physical bound for which the absorption efficiency is very large. In other words, the cloak opens the interesting possibility to achieve plasmonic resonances in the upper portion of the plot, above the point of maximal allowed absorption, for which the absorption efficiency is large and the scattering is suppressed. These resonant conditions are similar to the ones originally envisioned in [21], for which absorption and scattering cross-sections reach a local maximum at the same frequency, but with large ratios between the two. Plasmonic resonances with high absorption efficiency cannot be supported by a bare nanoparticle with similar size, as seen in Figure 4. The only solution to get a high level of absorption with a bare particle is to exploit the transparency condition, characterized by a scattering dip and nonresonant (flat) absorption cross-section, consistent with some of the concepts discussed in [41]. A first evident advantage of using a suitably designed cloak is then to open the high-efficiency region to plasmonic resonances, enabling resonant sensors with low visibility.

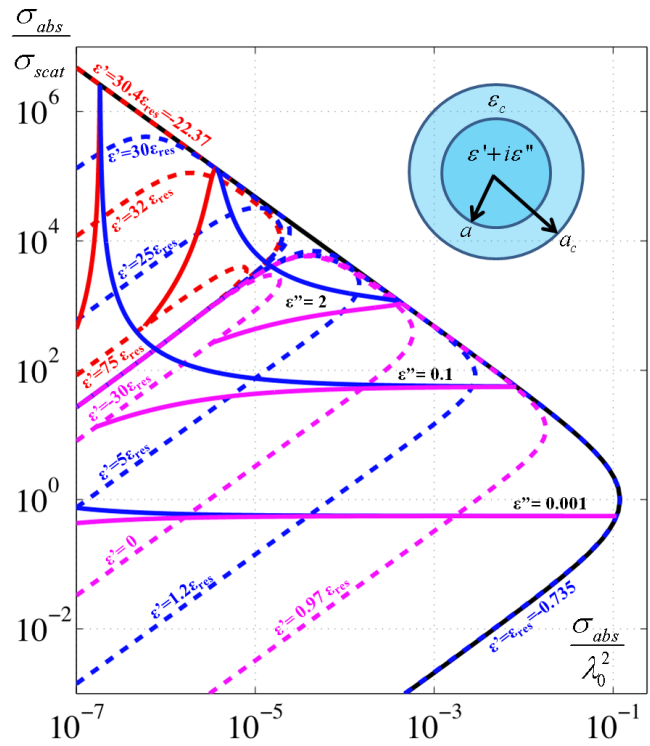

Figure 6. Same as Figure 4, for a core-shell nanoparticle of electrical size $k_0 a_c = 0.2$, with $\varepsilon_c = 0.105$ and $a = 0.9 a_c$.

Obviously the presence of the cloak can also provide more flexibility to choose the values of $\varepsilon'$ for the resonance and transparency conditions. In the example of Figure 6 these values are both negative: $\varepsilon_{resonance} = -0.7$ and $\varepsilon_{transparency} = -22.4$, which may be derived by studying the core-shell structure in the quasi-static limit, generalizing the analysis in the previous section. For a core-shell nanoparticle, the conditions to lie on the bound become

$$\varepsilon'_\pm = \frac{f(\varepsilon_c,\gamma) \pm \sqrt{729\varepsilon_c^2\gamma^6 - \varepsilon''^2 g(\varepsilon_c,\gamma)}}{g(\varepsilon_c,\gamma)} \qquad (23)$$

with

$$f(\varepsilon_c,\gamma) = -4\varepsilon_c^3(1+\gamma^3-2\gamma^6) - 4\varepsilon_c(-2+\gamma^3+\gamma^6) - \varepsilon_c^2(4+\gamma^3+4\gamma^6) \qquad (24)$$

and

$$g(\varepsilon_c,\gamma) = 2(2+\varepsilon_c+2(\varepsilon_c-1)\gamma^3)(\varepsilon_c-1+\gamma^3+2\varepsilon_c\gamma^3). \qquad (25)$$

As expected, Eq. (23) collapses to the bare nanosphere case Eq. (12) when we consider $\gamma = 1$. The condition to reach the bound now depends on $\varepsilon_c$ and $\gamma$: $729\varepsilon_c^2\gamma^6 - \varepsilon''^2 g(\varepsilon_c,\gamma) > 0$. In the low loss limit, the two solutions simplify into

$$\varepsilon'_\pm = \frac{\pm 27\varepsilon_c^2\gamma^3 + \varepsilon_c(-4(-2+\varepsilon_c+\varepsilon_c^2) - (4+\varepsilon_c+4\varepsilon_c^2)\gamma^3 + 4(-1+\varepsilon_c)(1+2\varepsilon_c)\gamma^6)}{2(2+\varepsilon_c+2(-1+\varepsilon_c)\gamma^3)(-1+\varepsilon_c+\gamma^3+2\varepsilon_c\gamma^3)} + O(\varepsilon''^2). \quad (26)$$

This condition, evaluated for $\varepsilon_c = 0.105$ and $\gamma = a/a_c = 0.9$, indeed provides $\varepsilon'_- = -0.72$ and $\varepsilon'_+ = -21.7$, in good agreement with the values numerically obtained in Figure 6. In addition, one can easily verify, following the same steps as in the previous section, that in the quasistatic and low-loss limits, the solution $\varepsilon'_-$ coincides with the plasmonic resonance condition of a core-shell particle, and $\varepsilon'_+$ with the cloaking condition [36].

To understand why the PEC asymptote is shifted up in the presence of a plasmonic cloak, we write the equivalent of Eq. (18) for a core-shell structure. In the limit $\varepsilon'' \to +\infty$, we obtain that the asymptote is a straight line of the form $\log(\sigma_{abs}/\sigma_{scat}) = a\log(\sigma_{abs}/\lambda_0^2) + b$. The slope $a = 1$, as before, and the intercept of the vertical axis, which determines the PEC asymptote, is given by

$$b = \log\frac{\pi(4(9+x^6)(-1+\gamma^3)^2 + (9+4x^6)(\varepsilon_c + 2\varepsilon_c\gamma^3)^2 + 4\varepsilon_c(-9+2x^6)(-1-\gamma^3+2\gamma^6))}{6x^6(-1+\varepsilon_c+\gamma^3+2\varepsilon_c\gamma^3)^2}. \quad (27)$$

When the cloak is designed to cancel the scattering of a PEC sphere of same outer radius as the core-shell nanoparticle, i.e., under the condition [9]

$$\gamma = \sqrt[3]{\frac{1-\varepsilon_c}{1+2\varepsilon_c}}, \quad (28)$$

we indeed obtain $b \to +\infty$, which elegantly confirms the behavior of the cloaked sensor. This proves that one can arbitrarily shift upwards the PEC asymptote in the plot by designing a cloak that cancels the scattering of a PEC sphere of same outer radius as the considered particle. This is an important design rule for resonant cloaked sensors, as it enables peculiar plasmonic resonances in the high absorption efficiency region. In Figure 6, for which $\gamma = 0.9$, we are very close to the ideal value $\gamma = \sqrt[3]{179/242} \approx 0.904$ predicted by (28), enabling a significant upward shift of the asymptote. Obviously the fundamental bounds derived in the previous section are still respected by cloaked particles, and the main effect of adding a cloak consists in providing more flexibility in tailoring the dynamic relation between absorption and scattering.

In Figure 7, we explore another cloak design, i.e., coating the core with an epsilon-near-zero (ENZ) shell $\varepsilon_c = 0.01$. From Eq. (26) we see that the effect of an ENZ cloak is to bring closer the resonance and transparency conditions, yielding $\varepsilon'_+ - \varepsilon'_- \to 0$ when $\varepsilon_c \to 0$, consistent with the Fano-like scattering signatures obtained with ENZ cloaks in [43]. This is also verified in the dynamic case, as seen in the figure, in which resonance and transparency points are brought very close to each other, $\varepsilon'_+ = -0.096$ and $\varepsilon'_- = -0.11$. Such designs may be of interest to enhance nonlinear effects for switching applications, exploiting the strong on/off dependency of the scattering cross-section and absorption efficiency, extending the concepts proposed in [45] to the case of sensors with tunable efficiency. Again, the physical bound discussed here is crucial to understand the limitations, complexity and potential of these nanoswitching devices.

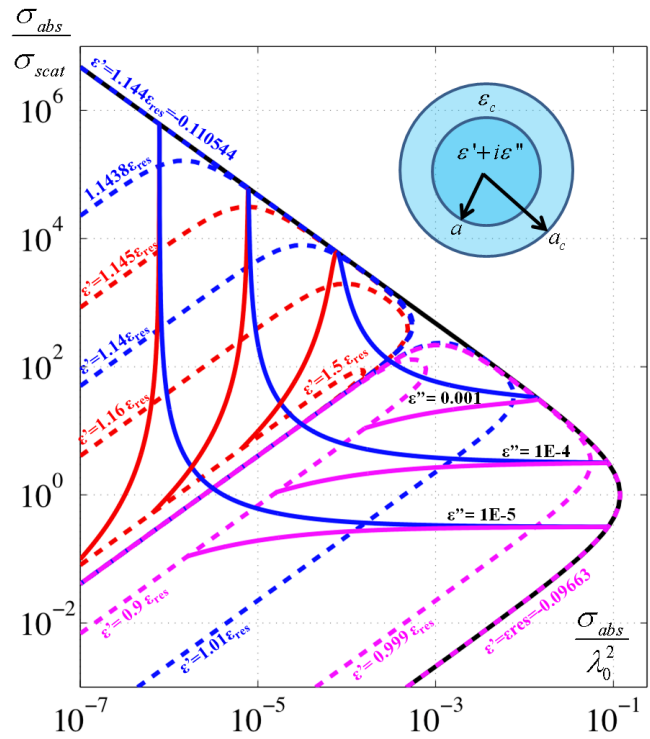

Figure 7. Same as Figure 6 but with a different cloak permittivity $\varepsilon_c = 0.01$.

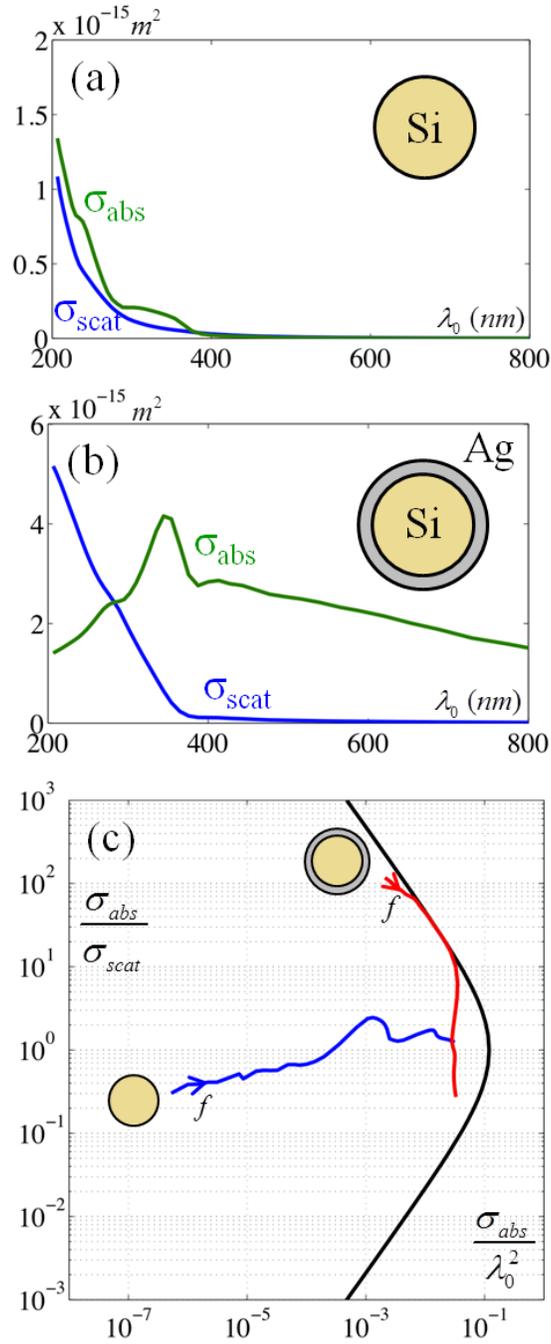

Figure 8. (a) Scattering and absorption spectrum for a realistic 40nm silicon particle at optical frequencies. (b) Scattering and absorption spectrum for the same particle cloaked by a 11nm plasmonic cloak made of silver. (c) Comparison of the scattering spectrum with the bound for the bare and cloaked case.

As a realistic scattering example, we consider the optical scattering from a silicon nanoparticle with radius $a = 20$ nm and compare it to the one obtained when the same sphere is surrounded by a plasmonic cloak made of silver. Material dispersion is taken from experimental data [46-47]. The scattering and absorption cross-sections of the bare sphere are reported in Figure 8(a) for incident wavelengths between 200 nm and 800 nm, and the contour obtained when sweeping frequency is shown in blue in Figure 8(c) and compared to the $TM_1$ bound. As evident from these plots, the bare silicon nanosphere starts absorbing significantly only in the UV range, due to increased electronic absorption processes at these energies. The absorption efficiency is close to unity throughout the optical range, making the nanosphere a rather inefficient absorber. In Figure 8(b), we consider the scattering spectrum of the same nanoparticle embedded in an 11 nm silver shell, and compared it to the bound in Figure 8(c) (red line). As evident from the figure, the cloak completely modifies the scattering properties of the structure, enabling it to access high absorption efficiency values ($>50$) in the visible range, and at the same time reaching the bound. These findings show that it is possible to largely manipulate and optimize the absorption and absorption efficiency of a nanosphere in the frequency range of interest by properly coating it, within the fundamental bounds derived in the previous section.

*C. Tunable antennas with optimal absorption efficiency*

As a final example to highlight the breadth of our findings, envision now a conventional radio-frequency sensor, consisting of an electrically small dipole antenna loaded by an impedance $Z_L = R_L - iX_L$, as considered in [41]. The associated scattering problem may be analytically solved, as shown in the Appendix, assuming without loss of generality that the antenna is aligned in the direction of polarization of the impinging field, by modeling the antenna as a dipole with polarizability [41]

$$\alpha^{-1} = -\frac{3\omega X_{in}}{l^2} \frac{X_{in} + X_L - iR_L}{4X_{in} + X_L - iR_L} + i\frac{k_0^3}{6\pi\varepsilon_0}, \tag{29}$$

where $X_{in}$ is the input reactance of the dipole antenna of half length $l$. Also in this case we choose a subwavelength geometry, so that the scattering is dominated by the dipolar contribution. The antenna length is $2l \approx \lambda_0/3 = 3$ cm, with a diameter of 600 μm, and we operate it at 3 GHz. By varying the load resistance and reactance, it is possible to tune the dipolar scattering and absorption of the object at will, similar to the previous plots for dielectric nanospheres, and span the whole admissible region in Figure 9, in which we calculate the TM[1] absorption efficiency and absorption cross-sections for this loaded dipole, for different values of $Z_L = R_L - iX_L$. In these calculations, we take the value of input reactance $X_{in} = -253.9 \Omega$, calculated for our geometry using formula (A13), and we restrict ourselves to $R_L \geq 0$, to ensure passivity. The solid contours are generated by keeping $R_L$ constant and sweeping $X_L$ from negative (capacitive) to positive (inductive) values. The dashed lines are generated by keeping $X_L$ constant and sweeping the load resistance through positive values. By intersecting these two sets of contour lines, it is possible to extract from the figure the absorption and absorption efficiency for any complex value of load impedance. Also here only contour lines with the same color intersect.

The figure shows similar features as in the previous examples, despite the completely different nature of the scatterer. The load resistance of the dipole antenna plays a role analogous to the imaginary part of the permittivity of the nanosphere, while the load inductance plays the role of the real part of permittivity, and similar considerations may be drawn as in the previous subsections.

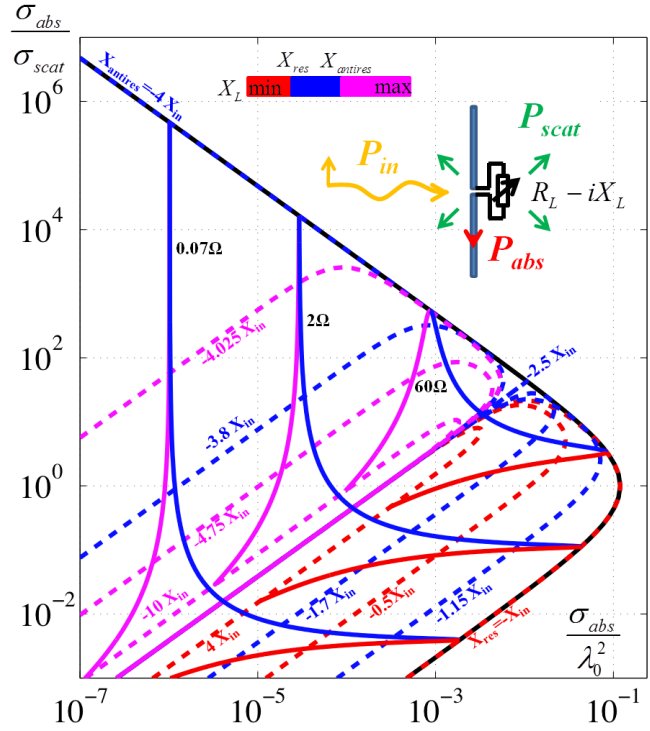

Figure 9. Similar to Figures 3-4, but for a loaded dipole antenna, varying its load impedance $R_L - iX_L$. The dashed contours are plotted varying $R_L$, for constant $X_L$ at the indicated value. The solid contours are plotted varying $X_L$, for constant $R_L$ at the indicated value (in black). The solid black line represents the fundamental limit for the first scattering harmonic.

We can explain the dynamics of the contours of Figure 9 by studying the system in the quasi-static limit. Repeating the steps detailed above in the case of the nanosphere, we find the two possible solutions to reach the physical bound

$$X_L^\pm = -\frac{1}{2}(5X_{in} \pm \sqrt{-4R_L^2 + 9X_{in}^2}). \tag{30}$$

Eq. (30) shows that the bound may only be reached for sufficiently low values of the load resistance, i.e. satisfying

$$0 \leq R_L \leq \frac{3}{2}|X_{in}|. \tag{31}$$

In the limit of a low load resistance $R_L \ll 3|X_{in}|/2$, the first condition simplifies into

$$X_L^+ = -4X_{in} + \frac{R_L^2}{3X_{in}} + O(R_L^3), \tag{32}$$

and the second, into

$$X_L^+ = -X_{in} - \frac{R_L^2}{3X_{in}} + O(R_L^3). \tag{33}$$

As in the previous examples, it is easy to check that condition (32) coincides with a scattering minimum in the low resistance limit, consistent with the findings in [41]. This is the transparency condition for a small loaded dipole. Conversely, Eq. (33) coincides with the condition to maximize the absorption for the given value of $R_L$ in the low resistance limit. This is the antenna resonance condition, analogous to the plasmonic resonance of the previous sections, for which the absorbed power is maximized. Also in this case there is a threshold, defined by (31), for the material losses $R_L$ beyond which we cannot reach the physical bound.

As seen in Figure 9, our quasi-static considerations are fully supported by the dynamic calculations: for fixed (small) load resistance (solid lines), the absorption cross-section is indeed maximized around the resonant condition $X_L = -X_{in}$, while the absorption efficiency is maximized around the transparency condition $X_L = -4X_{in}$, consistent with (32)-(33) and with the findings in [41]. The absolute maximum absorption is achieved under the conjugate matched condition $Z_L = Z_{in}^*$ [37], which provides unitary absorption efficiency. Similar to the case of nanoparticles, the contours of constant $X_L$ all converge to the same asymptote when $R_L \to +\infty$ or $X_L \to \pm\infty$. Indeed, in both cases we obtain the following equation for the asymptote

$$\log \frac{\sigma_{abs}}{\sigma_{scat}} = \log\left(\frac{2\pi}{3} + \frac{216\pi^3 X_{in}^2}{\eta_0^2 k_0^4 l^4}\right) + \log \frac{\sigma_{abs}}{\lambda_0^2}, \tag{34}$$

where $k_0$ and $\eta_0$ are respectively the wave number and characteristic impedance in free space, which represents the open-circuit asymptote, the analog of the PEC asymptote in the nanosphere case. Also here, this asymptote shifts up for shorter antennas. As evident in Figure 9, we also obtain a family of lossless asymptotes for constant $X_L$ in the limit $R_L \to 0$. These asymptotes are all parallel to the open-circuit asymptote with expression

$$\log \frac{\sigma_{abs}}{\sigma_{scat}} = \log\left(\frac{2\pi}{3} + \frac{216\pi^3 X_{in}^2 (X_{in} + X_L)^2}{\eta_0^2 k_0^4 l^4 (4X_{in} + X_L)^2}\right) + \log \frac{\sigma_{abs}}{\lambda_0^2} \tag{35}$$

and an upward shift when $X_L / X_{in} < (X_L^+ + X_L^-)/2$, and a downward shift when $X_L / X_{in} > (X_L^+ + X_L^-)/2$, with $(X_L^+ + X_L^-)/2 = -5/2$ in the quasi-static limit. The open-circuit asymptote separates the transparency and resonance regions, as discussed for the nanosphere scenario. This example demonstrates the generality of our analysis and of the derived physical bounds.

Like in the case of nanoparticles, the addition of a cloak may be used to effectively reduce the size of the antenna, shifting up all contours, consistent with the geometry originally proposed in [21]. These results clarify the potential of the cloaked sensor concept and the reach it may have in manipulating scattering and absorption within the bounds derived here. The cloak may allow achieving large absorption efficiencies in the resonant region, enabling the response discussed in [21], for which scattering and absorption both reach a local maximum at the design frequency, with large ratio between the two. In this case, the antenna may not be easily detectable when out of resonance (since it almost does not scatter), and it scatters the minimum for the chosen level of absorption, resulting in the best case scenario for passive cloaked sensors. Our findings may enable the design of optimized cloaks for tunable receiving antennas with high absorption efficiency and optimal minimum-scattering antennas. We will discuss these issues in further details in an upcoming study.

## IV. CONCLUSIONS

In this paper, we presented and discussed fundamental bounds on scattering and absorption of passive objects. These limitations, derived from passivity and power conservation, successfully quantify the minimum and maximum scattering for a given level of absorption, providing an important tool to qualitatively and quantitatively understand the limitations associated with cloaking absorptive objects. We applied our theory to a variety of examples, including optical scattering from dielectric nanospheres and core-shell nanoparticles, and microwave scattering from a loaded dipole antenna. We have explained the role of the cloak in cloaked sensor designs, showing that one can enable peculiar resonances for which both scattering and absorption reach a local maximum, but with a large ratio between them. The derived physical limitations provide a seminal basis in a wide range of situations. Our analysis may be readily applied to bigger objects for which several harmonics contribute to the scattering, as each of them follows the bounds described in this work, a concept that may be used to realize furtive super-absorbers [48]. Comparison with the bound presented here is a relevant figure of merit for any practical design of furtive sensors and absorbers. We believe that these fundamental bounds can be used to formulate a set of design rules to engineer optimal low-scattering sensors and absorbers in the optical regime, as well as minimum-scattering antennas at radio-frequencies.

ACKNOLEDGMENTS

This works has been supported by the AFOSR YIP award No. FA9550-11-1-0009 and the DTRA YIP award No. HDTRA1-12-1-0022.

APPENDIX

In this Appendix, we solve analytically the general problem of scattering of electromagnetic waves by a core-shell geometry loaded by a dipole of polarizability $\alpha$ placed at the center of a spherical coordinate system $(r, \theta, \varphi)$ centered with the core-shell, and oriented along $\hat{x}$. It is surrounded by the permittivity profile:

$$\varepsilon(r) = \begin{cases} \varepsilon & \text{if } 0 \leq r < a \\ \varepsilon_c & \text{if } a \leq r < a_c \\ \varepsilon_0 & \text{if } a_c \leq r < +\infty \end{cases} \quad (A1)$$

This general geometry includes all the possible scenarios analyzed in the present paper.

A plane wave $\vec{E}_{inc} = \hat{x}E_0 e^{ik_0 z}$ is incident upon the system, and we assume an $e^{-i\omega t}$ time-dependence. The total field, the sum of the scattered and incident field, may be expanded into spherical waves and decomposed into transverse electric (TE) and transverse magnetic (TM) fields, respectively associated with the radial magnetic and electric vector potentials [34,49]

$$\begin{cases} A_r = E_0 \dfrac{\cos\varphi}{\omega} \sum_{n=1}^{+\infty} i^n \dfrac{2n+1}{n(n+1)} \beta r \rho_n^{TM}(\beta r) P_n^1(\cos\theta) \\ F_r = E_0 \dfrac{\sin\varphi}{\omega\eta} \sum_{n=1}^{+\infty} i^n \dfrac{2n+1}{n(n+1)} \beta r \rho_n^{TE}(\beta r) P_n^1(\cos\theta) \end{cases}, \quad (A2)$$

where $P_n^m$ are Legendre polynomials and $\eta = \eta(r) = \sqrt{\mu_0/\varepsilon(r)}$ is the characteristic impedance of the medium, noted $\eta_0$, $\eta_c$ and $\eta$ in the outside medium, the cloak, and the inside domain, respectively. The wave number $\beta = \beta(r) = \omega\sqrt{\mu_0 \varepsilon(r)}$ will be noted $k_0$, $k_c$ and $k$ in the outside medium, the cloak, and the inside domain, respectively. The radial functions $\rho_n^{TE/TM}(\beta r)$ in (A2) are solutions of the radial equation, obtained by solving the spherical Helmholtz equation. Taking into account the excitation field, these functions can be sought, for TE waves, in the form

$$\rho_n^{TE}(r) = \begin{cases} a_n^{TE} j_n(kr) & \text{if } 0 \leq r < a \\ d_n^{TE} j_n(k_c r) + e_n^{TE} y_n(k_c r) & \text{if } a \leq r < a_c \\ j_n(k_0 r) + c_n^{TE} h_n^{(1)}(k_0 r) & \text{if } a_c \leq r < +\infty \end{cases}, \quad (A3)$$

where $j_n$ and $y_n$ are spherical Bessel functions of the first and second kind, and $h_n^{(1)}$ is the spherical Hankel function of the first kind. The TE radial functions (A3) are unaffected by the presence of the dipole, since dipolar radiation can be described as a $TM^1$ wave[36]. Therefore, the TM radial functions differ from the expression in (A3) by the addition of the field radiated by the dipole, yielding the form

$$\rho_n^{TM}(r) = \begin{cases} a_n^{TM} j_n(kr) + \delta_{1n} b_1^{TM} h_1^{(1)}(kr) & \text{if } 0 \leq r < a \\ d_n^{TM} j_n(k_c r) + e_n^{TM} y_n(k_c r) & \text{if } a \leq r < a_c \\ j_n(k_0 r) + c_n^{TM} h_n^{(1)}(k_0 r) & \text{if } a_c \leq r < +\infty \end{cases} \quad (A4)$$

where $\delta_{nm}$ is Kronecker's delta function and $b_1^{TM}$ depicts the strength of the radiation from the dipole antenna. The induced dipole moment at the center is linked with the local field by the polarizability $\alpha$ giving, after some calculations :

$$p = \alpha \left| \vec{E}_{loc} \right| = \alpha E_0 a_1^{TM}. \quad (A5)$$

On the other hand, the coefficient $b_1^{TM}$ may be related to the strength $p$ of the dipole, by expressing the usual dipolar radiation as a Mie series, i.e. [36]

$$b_1^{TM} = i \frac{k^3}{6\pi \varepsilon E_0} p. \quad (A6)$$

The coefficient $b_1^{TM}$ is now expressed as a function of $a_1^{TM}$ combining (A5) and (A6), then the result is inserted into (A4). Using (A2), (A3) and (A4), the tangential fields can be calculated. Enforcing the boundary conditions at $r = a$ and $r = a_c$ yields two linear systems,

$$\begin{pmatrix} -\dfrac{1}{\eta}\left[ j_n(ka) + i\delta_{1n}\alpha \dfrac{k^3}{6\pi\varepsilon_0} h_n^{(1)}(ka) \right] & \dfrac{j_n(k_c a)}{\eta_c} & \dfrac{y_n(k_c a)}{\eta_c} & 0 \\ -\dfrac{1}{ka}\left[ \hat{J}_n'(ka) + i\delta_{1n}\alpha \dfrac{k^3}{6\pi\varepsilon_0} \hat{H}_n^{(1)'}(ka) \right] & \dfrac{\hat{J}_n'(k_c a)}{k_c a} & \dfrac{\hat{Y}_n'(k_c a)}{k_c a} & 0 \\ 0 & \dfrac{j_n(k_c a_c)}{\eta_c} & \dfrac{y_n(k_c a_c)}{\eta_c} & -\dfrac{h_n^{(1)}(k_0 a_c)}{\eta_0} \\ 0 & \dfrac{\hat{J}_n'(k_c a_c)}{k_c a_c} & \dfrac{\hat{Y}_n'(k_c a_c)}{k_c a_c} & -\dfrac{\hat{H}_n^{(1)'}(k_0 a_c)}{k_0 a_c} \end{pmatrix} \begin{pmatrix} a_n^{TM} \\ d_n^{TM} \\ e_n^{TM} \\ c_n^{TM} \end{pmatrix} = \begin{pmatrix} 0 \\ 0 \\ \dfrac{j_n(k_0 a_c)}{\eta_0} \\ \dfrac{\hat{J}_n'(k_0 a_c)}{k_0 a_c} \end{pmatrix} \quad (A7)$$

for TM coefficients and

$$\begin{pmatrix} -\eta j_n(ka) & \eta_c j_n(k_c a) & \eta_c y_n(k_c a) & 0 \\ -\dfrac{\hat{J}_n'(ka)}{ka} & \dfrac{\hat{J}_n'(k_c a)}{k_c a} & \dfrac{\hat{Y}_n'(k_c a)}{k_c a} & 0 \\ 0 & \eta_c j_n(k_c a_c) & \eta_c y_n(k_c a_c) & -\eta_0 h_n^{(1)}(k_0 a_c) \\ 0 & \dfrac{\hat{J}_n'(k_c a_c)}{k_c a_c} & \dfrac{\hat{Y}_n'(k_c a_c)}{k_c a_c} & -\dfrac{\hat{H}_n^{(1)'}(k_0 a_c)}{k_0 a_c} \end{pmatrix} \begin{pmatrix} a_n^{TE} \\ d_n^{TE} \\ e_n^{TE} \\ c_n^{TE} \end{pmatrix} = \begin{pmatrix} 0 \\ 0 \\ \eta_0 j_n(k_0 a_c) \\ \dfrac{\hat{J}_n'(k_0 a_c)}{k_0 a_c} \end{pmatrix} \quad (A8)$$

for TE coefficients. We have used the following notation:

$$\hat{F}_n'(\beta r) = \dfrac{\partial}{\partial(\beta r)}\left[ \beta r f(\beta r) \right]. \quad (A9)$$

By solving the linear systems (A7) and (A8), the exact solution for the total field is obtained. The desired cross-sections can then be calculated using Eqs. (2)-(4).

Finally, we give the form of $\alpha$ for a loaded dipole antenna. For the fundamental physical bound to be respected, an accurate, power consistent expression of polarizability must be used. We assume the general form:

$$\alpha^{-1} = \alpha_S^{-1} + i\alpha_2^{-1}. \quad (A10)$$

If the scatterer is lossless, the static polarizability $\alpha_S^{-1}$ is real and $\alpha_2^{-1}$ is the radiation correction, taking care of power conservation. By plugging (A10) into (A7), assuming $\alpha_S^{-1} \in \mathbb{R}$ and imposing the lossless condition $\sigma_{scat} = \sigma_{ext}$, we obtain mathematically the necessary condition, imposed by power conservation:

$$\alpha_2^{-1} = -\frac{k^3}{6\pi\varepsilon}. \tag{A11}$$

The static polarizability $\alpha_S^{-1}$ is well-known for dipole antennas, and is expressed in the general case as [41]

$$\alpha_S^{-1} = -\frac{3\omega X_{in}}{l^2} \frac{X_{in} + iZ_L}{4X_{in} + iZ_L}, \tag{A12}$$

where $Z_L = R_L - iX_L$ is the complex impedance loading the antenna and $X_{in}$ is the negative imaginary part of the input impedance of the antenna, $Z_{in} = R_{in} - iX_{in}$, which can be approximated for small Hertzian dipoles with the following function of the antenna half-length $l$ and diameter $d$ [37]

$$X_{in} = -120 \frac{\ln\left(\frac{2l}{d}\right) - 1}{\tan(kl)}. \tag{A13}$$

The results presented in this Appendix cover all situations presented in this article, taking $\varepsilon_c = \varepsilon = 1$ for uncloaked antennas, $\alpha = 0$ for core-shell nanoparticles and $\alpha = 0$, $\varepsilon = 1$ for uncloaked spherical nanoparticles. The case of cloaked antennas is also included by the present calculation.